\documentclass{article}

\usepackage{arxiv}

\usepackage[utf8]{inputenc} % allow utf-8 input
\usepackage[T1]{fontenc}    % use 8-bit T1 fonts
\usepackage{hyperref}       % hyperlinks
\usepackage{url}            % simple URL typesetting
\usepackage{booktabs}       % professional-quality tables
\usepackage{amsfonts}       % blackboard math symbols
\usepackage{nicefrac}       % compact symbols for 1/2, etc.
\usepackage{microtype}      % microtypography
\usepackage{lipsum}
\usepackage{graphicx}
\usepackage{multirow}
\usepackage{amsmath}
\usepackage{tabularx}
\usepackage{array}
\usepackage{ragged2e}
\title{AI Education in a Mirror: Challenges Faced by Academic and Industry Experts}

\author{
  Mahir Akgun \\
  College of Information Sciences and Technology\\
  Pennsylvania State University\\
  University Park, PA 16802 \\
  \texttt{makgun@psu.edu} \\
   \And
 Hadi Hosseini \\
   College of Information Sciences and Technology\\
  Pennsylvania State University\\
  University Park, PA 16802 \\
  \texttt{hadi@psu.edu} 
  \\
}

\begin{document}
\maketitle
\begin{abstract}
As Artificial Intelligence (AI) technologies continue to evolve, the gap between academic AI education and real-world industry challenges remains an important area of investigation. This study provides preliminary insights into challenges AI professionals encounter in both academia and industry, based on semi-structured interviews with 14 AI experts—eight from industry and six from academia. We identify key challenges related to data quality and availability, model scalability, practical constraints, user behavior, and explainability. While both groups experience data and model adaptation difficulties, industry professionals more frequently highlight deployment constraints, resource limitations, and external dependencies, whereas academics emphasize theoretical adaptation and standardization issues. These exploratory findings suggest that AI curricula could better integrate real-world complexities, software engineering principles, and interdisciplinary learning, while recognizing the broader educational goals of building foundational and ethical reasoning skills. 
\end{abstract}

% keywords can be removed

\section{Introduction}
Artificial Intelligence (AI) is increasingly becoming a pivotal technology across various industries, driving innovation and solving complex problems. As AI systems become more integral to everyday life, ensuring that the next generation of AI professionals is well-equipped with both theoretical knowledge and practical skills is essential. Undergraduate AI education, therefore, plays a crucial role in preparing students to meet the demands of this rapidly evolving field and addressing the critical gaps that exist between academic training and industry requirements.

Considerable efforts have been made in AI curriculum development. Since 2018, the AI4K12 Initiative has been creating national guidelines for AI education in K-12 schools, focusing on the '5 Big Ideas in AI' \cite{AI4K12}. These guidelines outline the essential AI concepts and skills students should master at each grade level, providing a framework for curriculum developers and standards writers. Similarly, the ACM/IEEE-CS/AAAI's CS2023 guidelines emphasize core AI topics, ethical considerations, and interdisciplinary applications \cite{CS2023}. However, the fast-paced advancements in AI technology pose considerable challenges in maintaining a comprehensive and relevant curriculum. For instance, the rapid evolution of AI necessitates guidelines that remain relevant over extended periods. The development of the CS2023 guidelines exemplifies this challenge, as new techniques such as generative networks and large language models became widely used within just 18 months between the first draft and the final version \cite{eaton2024artificial}. Until the next revision, AI educators will need to determine which advancements to incorporate to ensure the curriculum remains up-to-date \cite{eaton2024artificial}. To address such ongoing changes, various initiatives have emerged to keep the curriculum dynamic and inclusive. These initiatives empower educators to continuously gauge and integrate recent advancements in AI, ensuring the curriculum reflects the latest developments. For instance, some initiatives integrate AI with STEM to foster project-based learning, passion, play, and peer learning in AI-related activities \cite{sakulkueakulsuk2018kids}. Other approaches have adopted layered designs to cover the multifaceted nature of AI, including key concepts and applications \cite{chiu2021creation}.

To ensure that undergraduate AI education not only keeps pace with technological advancements but also meets ever-changing industry needs, it is essential to understand the real-world challenges faced by AI professionals. Real-world AI problems often involve complex issues such as data scarcity, overcoming unrealistic assumptions, and navigating constraints defined by diverse stakeholders. These challenges provide critical learning opportunities for students, enabling them to develop a deeper understanding of the complexities and nuances of AI work. By identifying and analyzing these characteristics, educators can design undergraduate curricula that more accurately reflect the realities of the field, thus bridging the gap between academic learning and industry practice. This approach not only ensures that students are better prepared to tackle multifaceted challenges in their careers but also promotes the development of essential skills such as critical thinking, problem-solving, and ethical decision-making.

This study offers an exploratory examination of the challenges faced by AI experts in academia and industry, providing preliminary insights into potential gaps in AI education. Through semi-structured interviews with a limited but diverse group of experts, we identify emerging patterns and propose potential directions for better aligning AI education with real-world complexities, while acknowledging broader educational missions beyond immediate industry requirements.

\section{Prior Work}
\subsection{AI Education Challenges }
Artificial Intelligence (AI) education has undergone significant evolution in the past decade, but numerous challenges persist in aligning curricula with the rapidly changing demands of both academia and industry. One major difficulty highlighted by prior research is the translation of theoretical AI concepts into practical problem-solving skills \cite{sulmont2019can,allen2021toward}. To address this, AI pedagogy has increasingly explored methods such as project-based learning (PBL), interactive simulations, and real-world case studies, which help bridge the gap between theory and practice while enhancing student engagement \cite{denero2010teaching}.

However, students often face conceptual barriers that hinder their ability to grasp foundational AI concepts. These challenges include difficulties in understanding mathematical foundations, debugging AI models, and applying algorithmic decision-making in real-world contexts \cite{allen2021toward, skripchuk2022identifying}. In the realm of machine learning (ML), these barriers manifest as misconceptions about model behavior, such as overfitting, generalization errors, and incorrect hyperparameter tuning \cite{sulmont2019can} and misunderstanding of model-data relationships \cite{zimmermann2023common}. Thus, ML-specific challenges are a subset of broader AI education difficulties, underscoring the need for tailored educational strategies to address both general AI and specialized ML topics.

Pedagogical strategies such as scaffolding, active learning, and interdisciplinary integration have been proposed to alleviate these issues \cite{allen2021toward}. These methods aim to equip students not only with technical proficiency in AI and ML but also with a broader, contextual understanding of how AI systems operate in diverse, real-world environments \cite{sulmont2019can, allen2021toward}. Moreover, curriculum designs that incorporate active learning strategies, including problem-based learning and interactive engagement and real-time feedback, can help students address these errors more effectively, allowing them to improve their practical AI skills while reinforcing theoretical concepts.

\subsection{Software Engineering Perspectives in AI Education}
The interdisciplinary nature of AI training has led to increasing intersections between AI and software engineering (SE). Traditional software engineering principles often do not directly translate to AI-based systems, which rely on data-driven development rather than deterministic programming \cite{amershi2019software}. Research in software engineering for AI highlights challenges such as model versioning, continuous integration and deployment (CI/CD) for machine learning, explainability, and maintaining AI models in production environments \cite{baier2019challenges, paleyes2022challenges}. Studies have explored how AI education can incorporate software engineering methodologies, including MLOps, testing strategies for AI models, and modular software design principles that facilitate AI integration into complex systems \cite{bublin2021educating,chenoweth2023teaching}. Addressing these challenges requires AI curricula to equip students with hybrid skill sets that span machine learning, software engineering, and systems thinking \cite{amershi2019software}.

\subsection{Industry-Academia Gaps in AI Education}
The disconnect between AI education and industry expectations has been a recurring theme in computing education research. Studies have analyzed how AI graduates often lack exposure to real-world deployment challenges, such as data drift, model monitoring, scalability, and ethical considerations \cite{baier2019challenges, paleyes2022challenges}. Research in this area suggests that AI curricula should integrate interdisciplinary perspectives, including regulatory compliance, human-centered AI design, and ethical AI principles, to prepare students for diverse career paths \cite{paleyes2022challenges}. Additionally, prior studies highlight the need for experiential learning, where students work on industry-relevant AI problems that reflect the complexity of real-world applications \cite{amershi2019software, chenoweth2023teaching}. Approaches such as university-industry collaborations, capstone projects, and AI competitions have been explored as potential solutions to bridge this gap and better align academic AI training with professional requirements \cite{bublin2021educating,denero2010teaching}.
\section{Method}

This study employed semi-structured interviews with two distinct groups of AI experts from academia and industry. Participants answered a series of pre-defined questionnaires and open-ended questions aimed at uncovering the most challenging problems they face in their work. The responses were analyzed qualitatively using inductive content analysis, which facilitated the identification of key themes and labels characterizing common AI challenges. This analysis provided a foundation for comparing the challenges encountered by professionals and faculty. 

\subsection{Participants}
%Fourteen AI experts have been recruited for the study, comprising industry professionals and academics. 
To identify experts, we considered individuals with a degree in Computer Science, Information Sciences, Engineering, or related fields, and at least ten years of relevant experience. The criterion of ten years was chosen based on the widely accepted notion that extensive experience, often described as "10 years or 10,000 hours of deliberate practice," is indicative of expertise in a given field \cite{gladwell2008outliers}.

This study brought together fourteen experts from both industry and academia to provide comprehensive insights into the challenges facing AI development and education. The participant pool included eight AI industry practitioners working across diverse sectors (P1-P8) and six tenured faculty members with significant experience in AI research and teaching (A1-A6).

The industry participants represent a diverse range of organizations, including streaming services, social media platforms, e-commerce companies, pharmaceutical technology firms, and supply chain optimization companies. These practitioners offer perspectives from the front lines of applied AI development and implementation.

The academic experts, all tenured faculty members, teach both graduate and undergraduate courses while conducting research across various AI-related areas, including language models, data mining, machine learning, multi-agent systems, and computational social science.

Table 1 presents all participants, with the specialization field reflecting participants' self-reported areas of expertise. This diverse group of experts provides a well-rounded view of the practical challenges and educational considerations in the rapidly evolving field of artificial intelligence.                       

\begin{table}
\centering
\footnotesize
\setlength{\tabcolsep}{4pt}
\caption{Study Participants}
\label{tab:participants}
\begin{tabular}{l>{\centering\arraybackslash}p{1.5cm}>{\centering\arraybackslash}p{3cm}p{9cm}}
\toprule
\multicolumn{1}{c}{\textbf{ID}} & \multicolumn{1}{c}{\textbf{Sector}} & \multicolumn{1}{c}{\textbf{Organization/Position}} & \multicolumn{1}{c}{\textbf{Specialization}} \\
\midrule
P1 & Industry & Streaming services & AI model development, ML infrastructure, large-scale data systems \\
P2 & Industry & Social media platform & AI applications, scalable data solutions, user engagement \\
P3 & Industry & Social networking & Algorithm development, recommendation systems, UX improvement \\
P4 & Industry & Cloud/e-commerce tech & Deep learning, data processing, cross-product AI solutions \\
P5 & Industry & Tech \& supply chain & Algorithm optimization, model deployment, data-driven decisions \\
P6 & Industry & Pharmaceutical tech & Healthcare AI, predictive drug discovery models, patient data analysis \\
P7 & Industry/ Academia & Former industry professional & AI models, data science, ML infrastructure \\
P8 & Industry & E-commerce & E-commerce solutions, supply chain optimization, customer experience \\
A1 & Academia & Tenured Faculty & LLMs, deep learning with grammars/automata/rules, information retrieval \\
A2 & Academia & Tenured Faculty & Data management/mining, text, multimedia, social/web data, security applications \\
A3 & Academia & Tenured Faculty & Data mining, machine learning, social media mining \\
A4 & Academia & Tenured Faculty & AI, multi-agent systems, computational game theory, applied ML \\
A5 & Academia & Tenured Faculty & Computational/statistical ML, design and analysis of randomized algorithms \\
A6 & Academia & Tenured Faculty & NLP, privacy, security, computational social science \\
\bottomrule
\end{tabular}
\end{table}
\subsection{Research Design}
The study is structured to elicit the major challenges from experts in the field through semi-structured interviews lasting about 60 minutes. Semi-structured interviews involve a verbal interchange where the interviewer asks prepared questions while allowing the conversation to unfold naturally. This format enables participants to delve into issues they find important, providing richer and more nuanced insights. Unlike structured interviews, which strictly follow a set of predetermined questions, semi-structured interviews offer flexibility, encouraging a deeper exploration of topics.

We chose semi-structured interviews with open-ended questions for several reasons. Open-ended questions allow respondents to express their thoughts and experiences in their own words, leading to more detailed and contextually rich data. This is particularly important when exploring complex and subjective topics like the challenges faced by AI experts. Open-ended questions help avoid the limitations of closed-ended questions, which can restrict responses to predefined options and potentially overlook valuable insights.

The open-ended nature of our questions facilitates a conversational flow, encouraging participants to narrate their experiences and reflect on various aspects of their work. This approach aligns with our goal of understanding the diverse and multifaceted challenges encountered by AI professionals and academics.

The selection of the interview questions was driven by the goal of understanding not only the technical challenges faced by AI professionals but also the contextual factors that make certain problems particularly difficult or unique. The questions were designed to prompt participants to reflect on both specific experiences and the broader characteristics that distinguish routine tasks from particularly complex ones. This approach allowed us to explore the multi-dimensional nature of challenges in AI and identify patterns that might not be immediately apparent in more straightforward inquiries.
The questions posed in the interviews were:

\begin{itemize}
  \item \textit{Can you tell us about two-three most interesting or most challenging problems/cases you encountered in the past in your career? } \\
    This question was chosen to gather concrete, real-world examples of challenges that AI professionals have faced. By asking for specific problems or cases, the aim was to obtain responses grounded in practical experiences, ensuring that the discussion would focus on real-world challenges rather than theoretical concerns. This open-ended question encourages participants to select cases they found particularly memorable, offering insight into both the complexity and relevance of the challenges in their respective roles.
  
  \item \textit{Why do you think these cases/problems were especially interesting or challenging?}  \\
  This question delves into the reasoning behind the participant’s selection of cases, allowing them to reflect on the factors that made these problems stand out. By focusing on why these problems were challenging, we aimed to identify the key factors that make AI problems difficult, whether they are related to data, algorithmic complexity, or external constraints. This question helps to uncover the underlying causes of challenges, providing deeper insights into the nature of the problem-solving process.

  \item \textit{Are there any characteristics of these cases that are common with respect to the challenge? If yes, what are they? } \\
  This question explores patterns across the selected cases, seeking to identify common traits that may reveal larger trends or themes. By asking about shared characteristics, we aimed to surface systematic factors that contribute to challenges in AI, such as the nature of the data, the complexity of the model, or the constraints of the operational environment. This allowed us to identify not just isolated incidents, but recurring themes that characterize the types of problems that are most frequently encountered in the field.
  \item \textit{Are there any other characteristics that can be used to define “challenging/tough” problems/cases?} \\
  This question was included to broaden the scope of understanding by exploring any additional factors that define what makes a problem "challenging" or "tough." It was designed to capture a more generalizable perspective, allowing participants to reflect on broader, beyond the immediate context of their examples. It encouraged participants to think about qualitative factors such as team dynamics, resource limitations, or the level of uncertainty.
  
  \item \textit{What makes these tough/challenging problems/cases different from typical/routine problems/cases?} \\
  Finally, this question aimed to explicitly differentiate between routine tasks and exceptional challenges. By asking participants to compare and contrast the two, we were able to draw a clear distinction between common problems and those that require extraordinary problem-solving skills, greater resources, or innovative thinking. The response to this question provided crucial context for understanding the extraordinary nature of the cases highlighted in earlier questions, helping to identify the factors that truly set these problems apart from more typical ones.
\end{itemize}

\subsection{Data Analysis }
Before beginning analyses, the complete audio recordings of interviews were transcribed verbatim. Multiple readings of the transcripts were carried out to categorize the data into discrete episodes with respect to problems/cases that AI experts find challenging. 

The analysis focused on the challenges that experts have when developing and deploying AI solutions. The line-by-line reading was used as the analytical process of separating the transcribed data into constituent qualitative elements, but we also concentrated on portions of the data that were qualitatively meaningful units for signifying the challenges we aimed to identify in the study. A meaningful unit may be a line, a sentence, a paragraph, or any other entity, so we did not use a single entity as a unit of analysis in this study.  

The qualitative data were analyzed using a four-phase coding process to ensure consistency, transparency, and reliability in identifying key themes and patterns across responses.

\begin{itemize}
    \item Phase 1 – Independent Line-by-Line Coding\\
    Two researchers independently coded the interview transcripts in a line-by-line manner during the first pass. This inductive approach allowed both researchers to remain open to the emergence of new themes without being influenced by predefined categories. Each researcher reviewed the text carefully, assigning initial codes based on the content of the responses. This phase was entirely inductive, as the researchers aimed to identify patterns and categories based on the data itself, rather than applying pre-existing theoretical constructs.
    \item Phase 2 – Code Comparison and Rationalization\\
    After the initial coding, the researchers met to compare their codes and discuss the rationale for each assigned code. They identified instances where different terms were used to represent the same underlying concept. For example, when the two researchers used different codes to categorize similar themes, they engaged in a discussion to determine which term would best represent the case. This led to the refinement of the coding scheme, with the researchers selecting a single code to use consistently in subsequent analysis. This phase ensured that the coding terms were aligned and that both researchers had a shared understanding of how to apply the codes consistently.
    \item Phase 3 – Resolution of Discrepancies and Consensus\\
    In the third pass, the researchers focused specifically on any discrepancies between their initial codes. These discrepancies were discussed in detail, with each researcher providing the rationale for their choices. Through this collaborative process, they reached a consensus on the appropriate coding for each instance. This process ensured that all discrepancies were resolved, and the final dataset reflected a unified coding approach. Importantly, this phase guaranteed 100\% inter-rater reliability, as both researchers were in agreement on every code used.
    \item Phase 4 – Grouping Codes into Categories\\
    In the final step of the analysis, the researchers worked together to group related codes into broader categories that represented larger themes. This step was crucial for organizing the data in a meaningful way, allowing the researchers to synthesize the codes into categories that captured the key challenges and insights shared by the participants. These categories were developed through collaborative discussions, where the researchers examined the relationships between codes and identified overarching patterns in the data. This final categorization was the result of joint efforts, ensuring that the themes were well-grounded in the data and reflected the participants' perspectives.
\end{itemize}

\section{Results}

The analysis of interviews with AI experts from both industry and academia revealed a comprehensive set of characteristics that define the challenges they encounter in their work. These characteristics were categorized into distinct themes based on the qualitative data. Each theme represents a specific aspect of the challenges faced by AI professionals when developing and deploying AI solutions (see Table 2). 

\begingroup
\scriptsize % Smaller font size for the entire table
\setlength{\tabcolsep}{3pt} % Reduce column padding for better fit

% Use \FloatBarrier to prevent the table from floating past this point
%\FloatBarrier
\begin{table}  % H forces the table to be placed exactly here
\centering
\caption{Themes, codes, and their descriptions}
\label{tab:themes}
\begin{tabular}{p{3cm}p{3.5cm}p{8cm}}
\toprule
\textbf{Themes} & \textbf{Codes} & \textbf{Description} \\
\midrule

Data-Related Challenges & Imbalanced Data & Situations where the dataset has a significantly uneven distribution of classes affecting model performance. \\
\cmidrule{2-3}
& Lack of Good Data & Issues arising from the absence of high-quality domain-specific data necessary for robust model building. \\
\cmidrule{2-3}
& Limited Data & Challenges related to the availability of insufficient data to train models effectively. \\
\cmidrule{2-3}
& Low-Quality Feedback & Scenarios where feedback from users is inconsistent or not representative of actual performance. \\
\midrule

Model Adaptation and Scalability & Difficulty in Detecting New Kinds of Incidents & The challenge of identifying novel or evolving incidents that deviate from historical patterns. \\
\cmidrule{2-3}
& Handling Unpredictable Situations and Novel Contexts & This label covers scenarios where AI models encounter unforeseen behaviors or unfamiliar environments. \\
\cmidrule{2-3}
& Problems Involving Risk & Scenarios where decisions have significant potential consequences such as financial trading. \\
\cmidrule{2-3}
& Scalability Issues & The difficulty of scaling AI solutions from small-scale implementations to larger populations or settings. \\
\cmidrule{2-3}
& Irregular and Variable Data Structures & Dealing with irregular and variable data structures where relationships and connections between data points can vary greatly. \\
\cmidrule{2-3}
& Overcoming Unrealistic Theoretical Assumptions & The need to eliminate or adjust theoretical assumptions not feasible in real-world applications. \\
\cmidrule{2-3}
& Domain Knowledge Gaps & This label highlights difficulties AI practitioners face when they lack expertise in the specific domain where a model is applied. \\
\midrule

Practical Constraints and External Factors & Internal Data Influenced by External Factors & Situations where model accuracy is affected by external variables beyond the control of the dataset. \\
\cmidrule{2-3}
& Constraints Defined by Stakeholders & Limitations and requirements set by various stakeholders that influence AI system development. \\
\cmidrule{2-3}
& Constraints Shaped by Practical Settings & Practical limitations encountered in real-world environments differing from theoretical research settings. \\
\cmidrule{2-3}
& Resource and Infrastructure Constraints & This label encompasses limitations related to computational power, workforce availability, and financial resources. \\
\midrule

User Behavior and Interaction & Constraints Defined by User Action & Situations where user behavior introduces constraints that must be considered in model development. \\
\cmidrule{2-3}
& Making Incorrect Assumptions About Users & Scenarios where models are built based on incorrect assumptions about user behavior. \\
\cmidrule{2-3}
& Challenges in Understanding and Measuring User Impact & This label highlights the difficulty of predicting user responses to AI system outputs and defining appropriate long-term success metrics. \\
\midrule

Trust, Explainability, and Communication & Explainability and Trust Building & The necessity of making AI models transparent and understandable to build trust among stakeholders. \\
\cmidrule{2-3}
& Gap of Understanding & Communication barriers between technical and non-technical stakeholders leading to misaligned expectations. \\
\cmidrule{2-3}
& Overcoming Domain Expertise Resistance & Challenges in overcoming resistance from domain experts who may distrust AI models. \\
\bottomrule
\end{tabular}
\end{table}
%\FloatBarrier
\endgroup

\subsection{Identified Themes of AI Challenges}
This section presents the findings from the study, categorized into five overarching themes, each encapsulating specific challenges identified through participant interviews. These themes highlight the difficulties faced by AI practitioners in both academia and industry, illustrating how challenges manifest across different professional environments.

\subsubsection{Data-Related Challenges.}
 Data issues were frequently highlighted by experts, encompassing various aspects such as data quality, availability, and imbalance. These challenges directly impact the performance and reliability of AI models. Experts often face situations where the dataset has a significantly uneven distribution of classes (Imbalanced Data), making it difficult to train effective machine learning models because the model tends to be biased towards the majority class. One participant described this challenge:

 \begin{quote}
     “In terms of the most complex problems, not the ones that require the most sophisticated techniques, but I think in general, problems related to fraud, for example, are very challenging because the data is heavily imbalanced and you don't know what kind of fraud you would face in the future.” 
 \end{quote}
 In many enterprise settings, the user base may be small, resulting in a lack of adequate data (Limited Data). This scarcity makes it hard to fine-tune algorithms and make accurate predictions, which is particularly challenging when trying to deliver personalized or precise outputs.
 \begin{quote}
     “In enterprise settings, the user data is typically non-existent because the user base is very small, so the standard machine learning methods of query suggestions are not useful here, because you don't have any data to work with.” 
 \end{quote}
 
 Additionally, the absence of high-quality, domain-specific data necessary for building robust models (Lack of Good Data) was a recurring issue. This problem is exacerbated when there is no ground truth data or domain expertise to guide the model development and validation process. Furthermore, feedback from users that is inconsistent or not representative of actual system performance (Low-Quality Feedback) can mislead model improvement efforts, making it difficult for experts to assess the true efficacy of their systems.
 \begin{quote}
     “We don't actually end up getting reasonable feedback as to whether our price recommendation was good, or whether the user had different intentions and simply ignored the suggestion.” 
 \end{quote}

\subsubsection{Model Adaptation and Scalability.}
Adapting AI models to dynamic environments and ensuring their scalability emerged as a prominent concern among experts. Many AI models struggle to generalize beyond their training data, particularly when encountering new or unforeseen data patterns (Difficulty in Detecting New Kinds of Incidents). One participant highlighted this issue:
\begin{quote}
    “There are a lot of scams that happen these days, and not all scams follow the same patterns. Learning from the past doesn’t always help because newer fraud methods constantly emerge.” 
\end{quote}

Experts highlighted the need for models to handle novel contexts and unexpected scenarios (Handling Unpredictable Situations and Novel Contexts), where predefined rules or past experiences do not always provide sufficient guidance:
\begin{quote}
    “Unexpected surprises keep happening, and sometimes we aren't prepared for them. One bad actor can spoil the reputation of an entire marketplace.”
\end{quote}
In high-risk applications, such as finance or healthcare, the consequences of incorrect predictions are significant (Problems Involving Risk), making robust and fail-safe model design crucial. Moreover, scalability remains a persistent challenge (Scalability Issues), with many AI models failing to perform optimally when deployed at larger scales due to increasing computational costs and data variability. One expert noted that models often fail to scale effectively due to resource limitations and unpredictable system behavior:
\begin{quote}
    “During deployment, software behaves unpredictably, both in terms of input variations and the ways models interact with real-world environments.”
\end{quote}

Additional difficulties arise from processing non-standardized data structures (Irregular and Variable Data Structures) that demand flexible and adaptive algorithms. Finally, experts noted that theoretical assumptions often fail to align with real-world applications (Overcoming Unrealistic Theoretical Assumptions), leading to models that do not adequately reflect operational constraints. A lack of domain-specific knowledge (Domain Knowledge Gaps) further exacerbates these challenges, as AI solutions require contextual expertise to be effectively integrated into specialized fields.

\subsubsection{Practical Constraints and External Factors.}

Numerous constraints and external factors influenced the work of AI experts, highlighting the importance of considering real-world limitations in AI development. Various stakeholders set limitations and requirements (Constraints Defined by Stakeholders) that influence AI system development, often leading to conflicting expectations from senior executives, project managers, and end-users, complicating the development process.Stakeholder expectations play a crucial role in shaping AI development:
\begin{quote}
    “There is always ambiguity in problem settings and stakeholder expectations. People want solutions quickly, but the requirements keep evolving, making it difficult to define a stable approach.”
\end{quote}

Practical limitations encountered in real-world environments (Constraints Shaped by Practical Settings), such as unpredictable factors like traffic, weather, and operational constraints, differ from controlled or theoretical research settings. These constraints must be considered during model development to ensure applicability and effectiveness. Organizations face a broad range of limitations, such as shortages of people, money, and services (Resource and Infastructure Constraints), which necessitates efficient allocation of limited resources to maximize impact. Developing AI solutions that can operate on limited hardware resources (Low Compute Capabilities) is particularly relevant for organizations with limited budgets, posing a significant challenge for experts who need to ensure their models are both effective and resource-efficient. Model accuracy can also be affected by external variables beyond the control of the dataset (Internal Data Influenced by External Factors), such as economic changes, seasonal trends, or competitor actions. These factors introduce variability that impacts performance, making it challenging for experts to maintain model accuracy.
\begin{quote}
    “We can maximize recall in fraud detection, but this inevitably leads to false positives. The issue is that external factors continuously influence internal data, making fraud detection an ever-changing challenge.”
\end{quote}

\subsubsection{User Behavior and Interaction.}
The unpredictable nature of user behavior and interaction with AI systems posed significant challenges for the experts.
User behavior introduces constraints that must be considered in model development (Constraints Defined by User Action). For instance, users interacting with a chatbot in unexpected ways can require the system to handle off-topic or irrelevant queries effectively to maintain user engagement. Diverse and unpredictable user responses to system outputs can impact engagement and satisfaction, necessitating robust designs that can accommodate this unpredictability. Measuring the long-term impact of AI on user engagement and decision-making (Challenges in Understanding and Measuring User Impact) is another complex issue, as traditional evaluation metrics may not fully capture the evolving nature of AI-human interactions. 

Models built on incorrect assumptions about user behavior (Making Incorrect Assumptions About Users) often fail to meet actual needs and preferences, resulting in less effective models. Limited user research can lead to these incorrect assumptions, making it essential for experts to gather comprehensive and accurate user data:
\begin{quote}
    “We based our models on limited user interviews, assuming they represented the entire population. However, after deployment, we realized our assumptions were flawed, leading to unexpected failures.”
\end{quote}

\subsubsection{Trust, Explainability, and Communication.}
Building trust and ensuring clear communication between technical and non-technical stakeholders were crucial challenges for AI experts.
Making AI models transparent and understandable (Explainability and Trust Building) is essential for building trust among stakeholders. Explainable AI solutions help stakeholders understand how models arrive at their decisions, increasing their willingness to adopt and rely on these systems. However, achieving this transparency can be challenging, especially when dealing with complex models.

Communication barriers (Gap of Understanding) between technical and non-technical stakeholders can lead to misaligned expectations and solutions that do not fully address the intended issues. Ensuring effective communication and understanding is key to overcoming these barriers and aligning objectives:
\begin{quote}
    “One of the biggest challenges is interacting with non-technical stakeholders who struggle to articulate their problems in ways that AI researchers can interpret.”
\end{quote}
Additionally, professionals in specialized fields may resist AI-driven solutions due to concerns about reliability and lack of domain expertise in AI implementations (Overcoming Domain Expertise Resistance).
\begin{quote}
    “We had to convince domain experts that AI could enhance their work, but many resisted because they believed AI developers lacked the necessary field expertise.” 
\end{quote}

\subsection{Comparison of Challenges Between AI Professionals and Academics} 

To gain a deeper understanding of the distinct and overlapping challenges faced by AI experts in industry and academia, we conducted a comparative analysis of the qualitative codes extracted from interviews with AI professionals and faculty members. This analysis revealed commonalities and unique challenges encountered by each group, reflecting their different roles, priorities, and operational contexts. 

By examining these similarities and differences, we aim to highlight the multifaceted nature of AI challenges and the specific needs that must be addressed to bridge the gap between academic research and practical application. The following subsections provide a detailed summary of the codes observed in both groups, those unique to AI professionals and those specific to AI faculty members.

\subsubsection{Codes Observed in Both Groups.}
AI professionals and faculty members expressed several common challenges highlighting the universal issues in AI development and deployment. These shared challenges primarily revolve around data-related issues. Both groups emphasized the problem of imbalanced data, where datasets have a significantly uneven distribution of classes, complicating model training and performance. Additionally, limited data was a recurring theme, with experts from both domains expressing concerns about the availability of sufficient data to train effective models. This scarcity is particularly problematic in specific applications where large datasets are not easily obtainable. Furthermore, the lack of good data was another mutual concern, as professionals and academics recognized the difficulty in acquiring high-quality, domain-specific data crucial for building robust AI systems. These commonalities underscore the critical need for improved data collection and management practices across both academic and industry settings to enhance AI capabilities.

\subsubsection{Codes Extracted from AI Professionals.}
AI professionals highlighted a range of practical challenges that stem from their direct experiences in the field. One significant issue is low-quality feedback, where user feedback is often inconsistent or misleading, hindering the refinement of AI models. They also face difficulty in detecting new kinds of incidents and identifying emerging patterns, which requires continuous model adaptation to handle novel or evolving scenarios. Dealing with unexpected surprises in model behavior and problems involving risk, such as financial trading decisions, pose additional hurdles. Professionals must navigate unprecedented settings and manage internal data influenced by external factors, which can unpredictably affect model accuracy. Constraints defined by stakeholders and practical settings add layers of complexity, as do incorrect assumptions about users, constraints defined by user action, and the unpredictability of user reaction. Finally, professionals struggle with measuring long-term value for users and bridging the gap of understanding between technical and non-technical stakeholders, highlighting the multifaceted nature of their operational challenges.

\subsubsection{Codes Extracted from Academics.}

AI faculty members, on the other hand, emphasized challenges related to bridging theoretical research and real-world application. They pointed to the difficulties in adapting AI models for real-world implementation, which requires significant modifications from theoretical constructs. Overcoming unrealistic theoretical assumptions and domain expertise resistance are critical hurdles, as academic models often face skepticism from domain experts. Explainability and trust building are paramount, as faculty aim to make AI systems transparent and understandable to gain stakeholder trust. Additionally, faculty members highlighted low compute capabilities as a barrier, especially in resource-constrained environments. Scalability issues were also a major concern, reflecting the difficulties in expanding AI solutions from small-scale to larger applications. Resource limitations, such as resource constraints, lack of standardization in data structures, difficulty in defining universal principles, and dealing with irregular and variable data structures, further complicate their efforts. These challenges underscore the complexity of translating academic AI research into practical, scalable solutions that can be effectively implemented in various domains.
\section{Discussion}
The findings from our study provide valuable insights into the common challenges faced by AI experts. Below, we provide recommendations for educators to enhance AI curricula.

\subsection{Aligning AI Curricula with Real-World Challenges}
Traditional AI curricula emphasize algorithmic foundations, statistical modeling, and theoretical underpinnings, yet our study highlights significant real-world challenges that students may not encounter in a classroom setting. Issues such as imbalanced data, low-quality feedback, and resource constraints require AI practitioners to develop problem-solving skills beyond algorithmic implementation. Prior research has emphasized the importance of integrating pedagogical appraches that allow students to engage with real-world problems, gain hands-on experience and develop practical solutions. Allen et al. \cite{allen2021toward} highlight the importance of aligning teaching strategies with difficulties students in AI courses face when learning threshold concepts. Their study suggests best practices for teaching AI, including the use of practical examples and problem-based learning. Similarly, Sulmont et al. \cite{sulmont2019can} identify design decisions and model evaluation as the challenging aspects of AI education—areas that closely align with the challenges reported by industry professionals in our study. Furthermore, recent work by Skripchuk et al. \cite{skripchuk2022identifying} found that students often struggle when handling open-ended, real-world datasets, frequently making mistakes during data preprocessing and feature engineering stages. These findings reinforce the importance of preparing students to navigate the inherent messiness and unpredictability of real-world AI development environments.

\subsection{Integrating Software Engineering Perspectives into AI Education}

A key challenge identified in our study is that AI development often diverges from traditional software engineering (SE) practices. Unlike conventional software systems, AI models evolve based on new data inputs, making their maintenance, debugging, and deployment fundamentally different from rule-based software development. Yet, many of the challenges highlighted by AI practitioners in our study—such as scalability, resource constraints, unpredictability in deployment environments, and stakeholder constraints—mirror long-standing software engineering concerns \cite{baier2019challenges,chenoweth2023teaching}.

As Amershi et al. \cite{amershi2019software} emphasize, successful AI system development increasingly relies on software engineering practices such as continuous integration, monitoring, and iterative improvement, yet these are rarely emphasized in traditional AI coursework. Rather than treating AI model development as an entirely separate discipline, AI curricula can integrate core SE methodologies to equip students with hybrid skills necessary for real-world AI implementation. Prior research has emphasized that MLOps, testing strategies for ML models, and agile development frameworks are useful to keep AI models maintainable and scalable post-deployment \cite{chenoweth2023teaching}. These methodologies may provide structured approaches that have the potential to address several key AI challenges identified in our study, including: scalability Issues, model Adaptation and maintenance, and stakeholder-driven constraints.

\subsection{Bridging the Industry-
Academia Divide in AI Education}
A recurring theme in our study is the gap between AI education and industry expectations. While academia focuses on generalizable AI principles, industry professionals grapple with practical constraints such as resource limitations, evolving user behavior, and external dependencies.  Similar concerns were raised by Paleyes et al. \cite{paleyes2022challenges}, who found that operational and deployment challenges often receive insufficient attention in AI education, despite being critical barriers in industry practice. Multiple approaches can be considered to bridge this divide, including experiential learning opportunities such as university-industry collaborations and projects, which provide students with exposure to real-world problem-solving. Structured industry engagements, such as co-developed capstone projects, internships, and project-based coursework with external stakeholders, can provide students with early exposure to the practical realities of AI system development. Additionally, emphasizing interdisciplinary training — combining AI coursework with elements of ethics, and domain-specific expertise — can help students develop the flexibility needed to succeed across diverse AI applications. 

Importantly, while this paper emphasizes the value of aligning AI education with real-world professional challenges, we recognize that this is only one of several goals of higher education. Academic programs have historically pursued a variety of aims—including the cultivation of critical thinking, ethical reasoning, and theoretical inquiry—alongside workforce preparation. As Tedre et al. \cite{tedre2018changing} observe in their historical analysis, computing education has consistently served multiple overlapping purposes, and the emphasis on any one aim has shifted over time. Our focus on bridging the academia–industry gap reflects one important educational concern, not a prescription for the exclusion of others.

From this broader view, academic training serves not only to prepare students for specific professional roles, but also to foster foundational capabilities that support lifelong learning and critical engagement with evolving technologies. Denning \cite{denning2009profession} cautions against reducing computing education to algorithmic manipulation or symbolic problem-solving. He argues for a broader framing that treats computing as a professional practice, emphasizing principles such as computation, communication, coordination, automation, evaluation, and design. This perspective reinforces the importance of equipping students with reflective and transferable skills, which remain essential in an AI landscape marked by rapid technical and ethical change.

At the same time, AI curricula can benefit from a closer integration with practice-informed challenges. As Fincher and Petre \cite{fincher2004mapping} explain, educators often engage with expert practice not to replicate it wholesale, but to enhance students’ conceptual understanding and better support their transition into professional environments. This perspective reinforces the idea that academic curricula can selectively incorporate industry insights while maintaining their broader educational mission. In this sense, our study contributes to AI education by offering one perspective — grounded in practitioner experience — on how academic programs might evolve to address real-world complexity without compromising their foundational commitments.

\subsection{Integrating Real-World Problem Solving into AI Education}
To bridge the gap between academic learning and industry requirements in AI education, integrating project-based learning (PBL) can be highly effective. PBL allows students to engage with real-world problems, thereby developing deeper and more usable knowledge \cite{krajcik2022project}. Many of the skills essential for successful AI practice—such as critically evaluating models, making design decisions under uncertainty, and adapting solutions to dynamic conditions—are inherently higher-order cognitive tasks. Traditional lecture-based courses often fall short in cultivating these abilities, as they emphasize algorithmic mastery over application and critical analysis. Project-based learning, by contrast, provides an authentic context where students must actively apply theoretical knowledge, confront messy, open-ended challenges, and iteratively refine their solutions. Sulmont et al. \cite{sulmont2019can} also argue that students find higher-order machine learning tasks—such as model evaluation and design decision-making—especially challenging, further highlighting the need for project-based learning or similar learning approaches. By structuring AI education around authentic, complex projects, students can develop the analytical, evaluative, and design skills required to navigate real-world AI challenges.

By incorporating projects that simulate these challenges, students can gain hands-on experience and develop practical solutions. For instance, projects could involve designing AI systems that operate efficiently under limited computational resources, which mirrors the constraints often faced by nonprofits and other resource-limited organizations. This real-world application ensures that students understand the importance of optimizing algorithms and systems for environments where high-performance computing resources are not available.

Moreover, PBL encourages collaboration and social interaction, essential components in understanding and overcoming challenges related to user behavior and interaction with AI systems. As noted by Krajcik and Shin \cite{krajcik2022project}, social interactions in PBL environments help students construct shared understanding and engage in disciplinary practices. Projects that require students to work together to solve complex AI problems can mirror the collaborative nature of industry work, preparing them to navigate constraints defined by diverse stakeholders.

Engaging students in authentic tasks through PBL also helps them address the unpredictability of user reactions and the need for explainability in AI. Projects could involve developing AI systems for specific user groups, followed by testing and refining these systems based on user feedback. This iterative process not only enhances technical skills but also builds an appreciation for user-centered design and the ethical implications of AI technologies.

In addition to project-based work, incorporating structured failure analysis exercises into AI coursework could further strengthen students' critical thinking. By analyzing real-world cases where AI systems failed due to deployment challenges, ethical oversights, or model drift, students can develop a deeper understanding of the complex factors influencing AI system success in practice. Additionally, integrating basic MLOps concepts—such as continuous model integration, deployment monitoring, and iterative improvement—into AI courses would help students develop skills critical for maintaining real-world AI systems post-deployment.

\subsection{Limitations}
While our study provides valuable insights into the challenges faced by AI professionals and academics, it is important to acknowledge several limitations that may influence the interpretation and generalizability of our findings.

Despite efforts to include a range of perspectives, this study is limited by its relatively small sample size (n=14) and a participant pool primarily drawn from U.S.-based institutions and industries. As a result, the findings should be interpreted as exploratory and may not fully represent the global diversity of AI education and practice. The goal of this work is to offer preliminary insights rather than definitive conclusions, highlighting areas that warrant further investigation through broader, longitudinal, and cross-cultural studies.

The semi-structured interview methodology, while providing rich qualitative data, has inherent limitations. The open-ended nature of the interviews may lead to variability in the depth and breadth of responses, potentially influencing the themes and codes identified. Additionally, the reliance on self-reported data may introduce bias, as participants may emphasize certain challenges based on personal experiences or perceptions. Triangulating these findings with other data sources, such as surveys or observational studies, could provide a more comprehensive understanding of the challenges faced by AI professionals and academics.

\section{Conclusion}

Our study explored the challenges that AI professionals face in both industry and academia, highlighting key gaps between current AI education practices and the realities of professional AI development and deployment. While our findings are based on a relatively small, exploratory sample and should not be generalized without further validation, they offer preliminary insights into areas where AI curricula could evolve to better prepare students for real-world practice.

In light of these findings, we propose several strategies that could strengthen AI undergraduate education:

\begin{itemize}
    \item \textbf{Integrate Real-World Data Complexity into Coursework:} Incorporate projects and assignments using noisy, imbalanced, or incomplete datasets to expose students to practical data challenges.
    \item \textbf{Introduce Failure Analysis Exercises:} Embed structured analyses of real-world AI system failures into coursework to cultivate critical reflection on operational risks and system limitations.
    \item \textbf{Foster Interdisciplinary Collaboration:} Design learning experiences that involve working with domain experts and non-technical stakeholders, reflecting the cross-disciplinary nature of modern AI deployments.
    \item \textbf{Model User Behavior Variability in Design Projects:} Encourage students to anticipate and design for diverse, unpredictable user behaviors and evolving system requirements.
    \item \textbf{Promote Experiential Learning Opportunities:} Expand internships, industry-sponsored projects, and university-industry collaborations to offer students direct exposure to real-world AI development environments.
    \item \textbf{Strengthen Capstone Project Requirements:} Encourage capstone projects that simulate realistic resource constraints, dynamic conditions, and stakeholder negotiation processes.
\end{itemize}

These suggestions aim to complement the existing strengths of AI programs by better aligning technical instruction with the complexities and uncertainties encountered in real-world practice. By enhancing experiential, operational, and interdisciplinary training, AI curricula can foster a new generation of professionals who are technically adept, operationally resilient, and ethically aware.

We encourage future research to validate and extend these preliminary findings through broader, cross-institutional studies, ultimately refining educational practices to meet the evolving needs of the AI profession.

\bibliographystyle{unsrt}  


\bibliography{references}
\end{document}